\documentclass[a4paper,11pt]{article}

\usepackage{amsthm,amsmath,amssymb,amsfonts,pstricks,pst-node,graphics}

\usepackage{texdraw}

\DeclareMathAccent{\wtilde}{\mathord}{largesymbols}{"65}
\DeclareMathAccent{\what}{\mathord}{largesymbols}{"62}

\newcommand\cS{{\mathcal S}}
\newcommand\cF{{\mathcal F}}
\newcommand\cE{{\mathcal E}}
\newcommand\cT{{\mathcal T}}

\newcommand\Z{{\mathbb Z}}
\newcommand\N{{\mathbb N}}
\newcommand\C{{\mathbb C}}

\newcommand\cI{{\mathcal I}}
\newcommand\J{{\mathcal I}}
\newcommand\cH{{\mathcal H}}

\newcommand\cL{{\mathcal L}}
\newcommand\cM{{\mathcal M}}

\def\bbbc{{\mathbb C}}
\newcommand{\lieh}{{\mathfrak{h}}}
\newcommand{\gA}{{\mathfrak{A}}}
\newcommand{\gP}{{\mathfrak{P}}}

\newcommand{\gR}{{\mathfrak{R}}}
\newcommand{\gO}{{\mathfrak{O}}}
\newcommand{\gW}{{\mathfrak{W}}}
\renewcommand\d{{\rm d}}

\newcommand\bs{{\bf s}}
\newcommand\bt{{\bf t}}

\newtheorem{Def}{Definition}
\newtheorem{The}{Theorem}
\newtheorem{Pro}{Proposition}
\newtheorem{Lem}{Lemma}

\makeatletter
\def\wb{\accentset{{\cc@style\underline{\mskip10mu}}}}
\makeatother

\begin{document}
\title{Cosymmetries and Nijenhuis recursion operators for difference equations}
\author{Alexander V. Mikhailov$^{\star}$, Jing Ping Wang$ ^\dagger $
and Pavlos Xenitidis $^{\star}$\\
$\star$ Department of Applied Mathematics , University of Leeds, UK\\
$\dagger$ School of Mathematics, Statistics $\&$ Actuarial Science, University of Kent, UK }
\date{}
\maketitle

\begin{abstract}
In this paper we discuss the concept of cosymmetries and co--recursion operators for difference equations and
present a co--recursion operator for the Viallet equation. We also discover a new type of
factorisation for the recursion operators of difference equations. This factorisation enables us to 
give an elegant proof that the recursion operator given in {\it arXiv:1004.5346} 
is indeed a recursion operator for the Viallet equation. Moreover, we show that this operator is 
Nijenhuis and thus generates infinitely many commuting local symmetries. This recursion operator and its factorisation into
Hamiltonian and symplectic operators can be applied to Yamilov's discretisation of the Krichever-Novikov equation.
\end{abstract}

\section{Introduction}
The existence of hierarchies of infinitely many symmetries and/or conservation laws is one of the important characteristics 
of integrable equations. It has been successfully used in the classification of given families of partial differential and 
differential--difference equations \cite{mr80k:35060, mr86i:58070, mr89g:58092, mr89e:58062,
mr93b:58070, mr99g:35058, wang98, asy,Yami1,Yami}. 

The hierarchies of infinite symmetries can often be generated by recursion
operators mapping symmetries to symmetries \cite{AKNS74,mr58:25341,hssw05, Yami}.
Most of the known examples of recursion operators have the property of
being Nijenhuis operators. This property was independently
studied by Fuchssteiner--Fokas \cite{Fuc79,mr82g:58039,mr84j:58046} and Magri \cite{Mag80}, who named
such operators as hereditary symmetries. It guarantees that one can
generate infinitely many local commuting symmetry flows starting
from some seed symmetries. Furthermore, the conjugated recursion operators can generate infinitely many local
covariants (the variational derivatives of conserved densities). For Hamiltonian systems recursion operators can be 
factorised into symplectic and Hamiltonian operators. 

The most well known example is the famous Korteweg--de Vries (KdV) equation
$$ u_t\,=K=\,u_{xxx}\,+\,6\, u \,u_x\,,$$
which possesses a recursion operator
$$\Re\,=\,D_x^2\,+\,4\, u\, +\,2 \,u_x \,D_x^{-1}\,,$$
where \(D_x^{-1} \) stands for the inverse of \(D_x\). The operator $\Re$ satisfying 
$$D_{u_t-K} \Re=\Re D_{u_t-K}\ ,$$
where $D_{u_t-K}$ is the Fr{\'e}chet derivative along the KdV, is a Nijenhuis operator and generates the local commuting KdV hierarchy
\[
u_{t_j}\,=\,\Re^j (u_x),\qquad j\,=\,0, 1, 2,\cdots,
\]
the first member of which, i.e. a seed symmetry,  is $u_{t_0}=u_x$.

The operator conjugated  to the recursion operator $\Re$ for the KdV equation is
$$\Re^*=D_x^2+4 u -2 D_x^{-1} u_x$$
and satisfies the operator equation 
$$D_{u_t-K}^* \Re^*=\Re^* D_{u_t-K}^*\ .$$
It recursively produces infinitely many local covariants  $G_n=\Re^{*} (G_{n-1}),\ n\in\N$, with $G_0=1$, which are 
also cosymmetries of the KdV equation. Indeed, they satisfy $D_{u_t-K}^*(G_n)=0$. Thus, for the KdV equation the conjugate operator $\Re^*$ is a co-recursion 
operator mapping cosymmetries to cosymmetries.

The recursion operator $\Re$ can be represented as the composition of a symplectic operator $\cI$ and a Hamiltonian operator $\cH_1$, i.e.
\[ 
 \Re=\cH_1\cdot\cI,\qquad \cI=D_x^{-1},\qquad  \cH_1=D_x^3+4 u D_x+2u_x.
\]
The KdV equation is a multi-Hamiltonian system with infinitely many Hamiltonian operators $\cH_n=\Re^n\cdot \cH_0,\ n\in\N$, where $\cH_0=\cI^{-1}=D_x$. It is not difficult to show that all Hamiltonian operators are weakly non-local. 
The concept of weak non-locality was discussed in detail in \cite{MaN01}. 

In our recent work \cite{mwx1},  we adapted the concept of recursion operators for  difference equations. In particular,  we claimed that the operator $\gR$ given in Theorem 5 of \cite{mwx1} (see also (\ref{weakR}) below), 
is a recursion operator for the Viallet equation \cite{Viallet}:
\begin{eqnarray}
Q &:=& a_1 u_{0,0} u_{1,0} u_{0,1} u_{1,1} \nonumber \\
& & + a_2 (u_{0,0} u_{1,0} u_{0,1} + u_{1,0} u_{0,1} u_{1,1}
 + u_{0,1} u_{1,1} u_{0,0} + u_{1,1} u_{0,0} u_{1,0}) \nonumber\\
&&  + a_3 (u_{0,0} u_{1,0}+u_{0,1} u_{1,1}) +a_4 (u_{1,0} u_{0,1} +
u_{0,0} u_{1,1})  \nonumber\\
&&  + a_5 (u_{0,0} u_{0,1} + u_{1,0} u_{1,1}) + a_6
(u_{0,0}+u_{1,0}+u_{0,1} +u_{1,1})
+ a_7 \,=\,0\,,\label{QV}
\end{eqnarray}
where $a_i$ are free complex parameters. Equation (\ref{QV}) is a quite general and useful example 
of an integrable difference equation.
By a point fractional-linear
transformation this equation with a generic choice of parameters can be
reduced to Adler's equation, also referred to as the Q4 equation in the ABS
classification \cite{ABS, ABS1}. All of the ABS equations can be obtained from
the Viallet equation by a simple specialisation of the parameters. Thus, the results obtained for  
the Viallet equation can be easily applied to all ABS equations.

In this paper, we introduce and discuss the concepts of cosymmetries and 
co-recursion operators for difference equations.  We  give a complete proof of Theorem 5 formulated in \cite{mwx1}, namely that the operator $\gR$ presented in the theorem is 
indeed a recursion operator of the Viallet equation (\ref{QV}) and it is a composition of Hamiltonian and symplectic 
operators. Moreover, this operator possesses the Nijenhuis property and thus
produces infinitely many commuting local symmetries starting from two seed symmetries. Equation (\ref{QV}) possesses 
an infinite hierarchy of cosymmetries which can be generated by a co-recursion operator. Here  we would like to stress 
that the cosymmetries of equation (\ref{QV}) do not coincide with its covariants and that the co-recursion operator is 
not conjugated to the recursion one. As a by-product, we have found a recursion operator and its factorisation into
Hamiltonian and symplectic operators for the differential-difference integrable equation known as Yamilov's discretisation of the Krichever-Novikov equation \cite{Yami}.  

In order to be self-contained we introduce our notations, sketch the framework and give basic definitions 
in Section \ref{Basic}. More detailed description of the framework and motivations for the definitions the reader 
can find in our previous paper \cite{mwx1}. In Section \ref{Basic1} we define the field of fractions $\cF_Q$ 
corresponding to a difference equation $Q=0$, the elimination map $\cE$ and the set of dynamical variables. 
In the next subsection we give the definitions of symmetries and cosymmetries illustrating by useful examples 
for equation (\ref{QV}).

In Section \ref{Co-recursion} we discuss difference and pseudo-difference operators. In particular we define weakly 
non-local pseudo-difference operators. They are difference analogues of weakly non-local pseudo-differential operators  
in the theory of Poisson brackets (see A.Ya. Maltsev and S.P. Novikov   \cite{MaN01}). We give
the definitions of recursion and co-recursion operators for difference equations and show that operator 
$\gR$ (\ref{weakR}) is a recursion operator for the Viallet equation (\ref{QV}). On the course of the proof we 
discover a new factorisation property of the recursion operator (\ref{weakR}). Namely, there is a constant 
$\mu$ (\ref{const}) such that the operator $\gR-\mu$ can be factorised over the field $\cF_Q$ into a difference and 
a weakly non-local pseudo-difference operator. Interestingly enough, the coefficients of $\gR$ are elements of the subfield $\cF_\bs\subset\cF_Q$ (of rational functions of variables $\{u_{n,0}\,|\, n\in\Z\}$, see Section \ref{Basic1}) while the coefficients of the factors are in $\cF_Q$ and cannot be restricted to  $\cF_\bs$. A co-recursion operator for equation (\ref{QV}) is given explicitly in Theorem \ref{th2}.

In Section \ref{locality} we discuss Hamiltonian and symplectic operators related to difference equations and show that 
the recursion operator (\ref{weakR}) is hereditary (Nijenhuis). The main result of this section is Theorem \ref{th3},
which states that symmetries of equation (\ref{QV}) generated by the recursion operator $\gR$ (\ref{weakR}) are all local and commuting, despite of $\gR$ being a weakly non-local pseudo-difference operator.  The definitions of Hamiltonian and symplectic operators, the hereditary property of a recursion operator as well as the proofs of assertions are not different from the continuous case if we reformulate the problem in terms of a variational complex.
It enables us easily to adapt the theory developed for the continuous case   \cite{GD79, mr94j:58081, wang98, wang09} to 
the difference one.

\section{Framework and basic definitions}\label{Basic}
To make the paper self-contained, we give the basic definitions of the elimination map, dynamical variables, symmetries
and cosymmetries of difference equations. Motivations for
these definitions have been discussed in detail in our paper \cite{mwx1}.
\subsection{Difference equations and dynamical variables}\label{Basic1}
Difference equations on $\Z^2$ can be regarded as a difference analogous of partial
differential equations. Let us denote by $u=u(n,m)$ a complex-valued function
$u:\Z^2\mapsto \C$, where $n$ and $m$ are ``independent variables'' and $u$ will play
the r\^ole of a ``dependent'' variable in a difference equation. Instead of partial
derivatives we have two commuting shift maps $\cS$ and $\cT$ defined as
\[\begin{array}{c}
 \cS: u\mapsto u_{1,0}=u(n+1,m),\\ \cT: u\mapsto u_{0,1}=u(n,m+1),\\ \cS^p\cT^q:
 u\mapsto u_{p,q}=u(n+p,m+q). \end{array}
\]
For uniformity of the notation, it is convenient to denote the ``unshifted'' function
$u$ as $u_{0,0}$.  In the theory of difference equations we shall treat $u_{p,q}$ as variables.
The set of all shifts of the variable $u$ will be denoted by  $U=\{u_{p,q}\,|\, (p,q)\in\Z^2\}$.
For a function $f=f(u_{p_1,q_1},\ldots ,u_{p_k,q_k})$ of  variables $u_{p,q}$
the action of the operators $\cS,\cT$  is defined as
\[
 \cS^i\cT^j (f)=f_{i,j}=f(u_{p_1+i,q_1+j},\ldots ,u_{p_k+i,q_k+j}).
\]

In this paper, we consider a quadrilateral difference equation of the form
 \begin{equation}\label{Qequation}
Q(u_{0,0},u_{1,0},u_{0,1},u_{1,1})=0\ ,
\end{equation}
where $Q(u_{0,0},u_{1,0},u_{0,1},u_{1,1})$ is an irreducible polynomial of the
``dependent variable'' $u=u_{0,0}$ and its shifts. Irreducibility means that 
 $Q$ cannot be factorised and presented as a product of two polynomials. It is
assumed that equation (\ref{Qequation}) is valid at every point $(n,m)\in\Z^2$ and
thus (\ref{Qequation}) represents the infinite set of equations
\begin{equation}\label{Qequation1}
 Q_{p,q}=Q(u_{p,q},u_{p+1,q},u_{p,q+1},u_{p+1,q+1})=0, \qquad (p,\ q)\in\Z^2\ .
\end{equation}
Moreover, we assume that $Q$ is an irreducible
affine-linear polynomial which depends non-trivially on all variables, i.e.
\begin{equation}\label{affinelinear}
 \partial_{u_{i,j}} Q\ne 0,\ \ \partial^2_{u_{i,j}} Q=0,\
\qquad i,j\in\{0,1\},\ Q\in\C[u_{0,0},u_{1,0},u_{0,1},u_{1,1}].
\end{equation}
Here and later in the text $ \partial_{u_{i,j}} f$ refers to the partial derivative of $f$ with respect to $ u_{i,j}$ which we also often denote as $f_{u_{i,j}}$.

Let $\C[U]$  be the ring of polynomials of the variables
$U$. Maps $\cS$ and
$\cT$ are automorphisms of $\C[U]$ and thus $\C[U]$ is a difference ring. We denote
$J_Q=\langle \{Q_{p,q}\}\rangle$ the ideal generated by the difference equation  and
all its shifts (\ref{Qequation1}). It is a prime difference
ideal and thus $\C[U]/J_Q$ is a difference quotient ring without zero divisors. The corresponding field of fractions
we denote as
\[ \cF_Q=\{[a]/[b]\,|\, a,b\in\C[U],\ b\not\in J_Q\}\,, \]
where $[a]$ denotes the class of equivalent polynomials  (two polynomials $f,g\in\C[U]$
are equivalent if $f-g\in J_Q$). For $a,b,c,d\in\C[U],\  b,d\not \in J_Q$, the two rational functions
$a/b$ and $c/d$ represent the same element of $\cF_Q$ if $ad-bc\in J_Q$.
The field $\cF_Q$ is the main object in our theory.

The fields of rational functions of variables
\[
U_\bs=\{u_{n,0}\,|\, n\in\Z\}, \quad U_\bt=\{u_{0,n}\,|\, n\in\Z\},\quad  U_0=U_\bs\cup U_\bt.
\]
 are denoted respectively as
\[
 \cF_\bs=\C(U_\bs),\qquad \cF_\bt=\C(U_\bt),\qquad  \cF_0=\C(U_0)\,.
\]

Since the ideal $J_Q$ is generated by  affine-linear
polynomials $Q_{p,q}$, we  can define the {\em elimination} map $\cE$ which simplifies computations modulo
the ideal drastically. One can uniquely resolve equation $Q=0$ with respect to each
variable
\begin{equation}
 \label{subsQ}
\begin{array}{ll}
u_{0,0}=F(u_{1,0},u_{0,1},u_{1,1}),\qquad &u_{1,0}=G(u_{0,0},u_{0,1},u_{1,1}), \\
u_{0,1}=H(u_{0,0},u_{1,0},u_{1,1}),\qquad &u_{1,1}=M(u_{0,0},u_{1,0},u_{0,1})
\end{array}
\end{equation}
and since  $Q$ is an affine linear polynomial, functions  $F,G,H$
and $M$ are rational functions of their arguments.
Equations (\ref{subsQ}) enable us  recursively and uniquely to
express any variable $u_{p,q}$ in terms of the variables $U_0=U_\bs\cup U_\bt$. 
\begin{Def}\label{Def1}
 For elements of $U$ the {\em elimination map} $\cE:U\mapsto\C(U_0)$ is defined recursively:
\begin{equation}\label{Emap}
\begin{array}{ll}
\forall p\in\Z,\ \qquad &\cE(u_{0,p})=u_{0,p},\qquad  \cE(u_{p,0})=u_{p,0}\, ,\\
{\rm if}\ p>0,q>0,\  \quad &\cE(u_{p,q})=M(\cE(u_{p-1,q-1}),\cE(u_{p,q-1}),\cE(u_{p-1,q}))\,,\\
{\rm if}\ p<0,q>0,\  \quad &\cE(u_{p,q})=H(\cE(u_{p,q-1}),\cE(u_{p+1,q-1}),\cE(u_{p+1,q}))\,,\\
{\rm if}\ p>0,q<0,\  \quad &\cE(u_{p,q})=G(\cE(u_{p-1,q}),\cE(u_{p-1,q+1}),\cE(u_{p,q+1}))\,,\\
 {\rm if}\ p<0,q<0,\ \quad &\cE(u_{p,q})=F(\cE(u_{p+1,q}),\cE(u_{p,q+1}),\cE(u_{p+1,q+1}))\, .
 \end{array}
\end{equation}
For  polynomials $f(u_{p_1,q_1},\ldots ,u_{p_k,q_k})\in\C[U]$ the   elimination map $\cE:\C[U] \mapsto \C(U_0)$ is defined as 
$$\cE:f(u_{p_1,q_1},\ldots ,u_{p_k,q_k})\mapsto f(\cE(u_{p_1,q_1}),\ldots ,\cE(u_{p_k,q_k}))\in\C(U_0).$$ 

\noindent
For rational functions $a/b,\
a,b\in\C[U], \ b\not\in J_Q$ the elimination map $\cE$ is defined as 
$$\cE:a/b \mapsto\cE(a)/\cE(b).$$
\end{Def}

It follows from the above definition that $\cE:J_Q\mapsto 0$.
The elimination map is the difference field isomorphism $\cE\,:\, \cF_Q\mapsto \cF_0$. Any element of 
$\cF_Q$ can be uniquely represented by a rational function of variables $U_0$.
Variables $U_0$ we shall call the {\em dynamical} variables.

\subsection{Symmetries and cosymmetries of difference equations}

Symmetries and conservation laws of difference equations have been discussed in detail in our paper \cite{mwx1}. 
In this section we recall the definition of a continuous symmetry of a difference equation and give the definition of a  cosymmetry. 
For difference equations
cosymmetries do not coincide with covariances (the variational derivatives of conserved densities) as in the 
case of evolutionary equations. Cosymmetries are new objects in the theory of difference equations, 
and they will play important r\^ole in our construction of recursion and co-recursion operators. 

\begin{Def} \label{DefSymmetry}
Let $Q=0$ be a difference equation. Then $K\in\cF_Q$ is called a {\em generator of an
infinitesimal symmetry} (or simply, a {\em symmetry}) of the difference equation if
\begin{equation}\label{sym_def}
D_Q(K)= 0.
\end{equation}
Here $D_Q$ is the Fr\'echet derivative of $Q$ defined as
\begin{equation}\label{frechet}
D_Q=\sum_{i,j\in\Z}Q_{u_{i,j}} \cS^i \cT^j\ ,\qquad Q_{u_{i,j}}=\partial_{u_{i,j}} Q.
\end{equation}
\end{Def}

For a quadrilateral equation the sum in
(\ref{frechet})
has only four terms. In (\ref{sym_def}) $D_Q(K)$ is equal to zero as an element of $\cF_Q$. The way to check this is to apply the elimination map and therefore $K$
is a symmetry if $\cE(D_Q(K))=0$. 
If the difference equation $Q=0$ admits symmetries, then they form a Lie algebra
denoted as $\gA_Q$ \cite{mwx1}.

Similarly, we define cosymmetries:
\begin{Def} \label{DefcoSymmetry}
Let $Q=0$ be a difference equation. Then $\omega\in\cF_Q$ is called a {\em cosymmetry}
(or  a {\em characteristic of a conservation law} ) of the difference equation if
\begin{equation}\label{sym_def1}
D_Q^*(\omega)= 0,
\end{equation}
where $D_Q^*$ is a formally adjoined operator of the Fr{\'e}chet derivative which is defined as
\begin{equation}\label{frechet1}
D_Q^*=\sum_{i,j\in\Z}\cS^{-i} \cT^{-j} Q_{u_{i,j}} .
\end{equation}
\end{Def}
Cosymmetries of equation $Q=0$ form a linear space over $\bbbc$ which we denote as $\gO_Q$.

We shall see that equation (\ref{QV}) has infinite hierarchies of symmetries and cosymmetries which can be naturally 
expressed in terms of the discriminants of $Q$
\begin{eqnarray}
h(u_{0,0},u_{1,0}) &:=&  Q\, \partial_{u_{0,1}} \partial_{u_{1,1}} Q\, -\,
\partial_{u_{0,1}} Q \, \partial_{u_{1,1}} Q \,; \label{h-polynomials-def1}\\
g(u_{1,0},u_{0,1}) &:=&  Q\, \partial_{u_{0,0}} \partial_{u_{1,1}} Q\, -\,
\partial_{u_{0,0}} Q \, \partial_{u_{1,1}} Q \,; \label{g-polynomials-def1}
\end{eqnarray}
and the function 
\begin{equation}
 w=\frac{1}{u_{1,0}-u_{-1,0}}.\label{w}
\end{equation}
For any $f\in\cF_Q$ we shall use the abbreviated notation $f_k=\cS^k (f)$. In particular 
\[ 
 h_k=\cS^k h(u_{0,0},u_{1,0}),\quad g_k=\cS^k g(u_{1,0},u_{0,1}),\quad w_k=\cS^k w,
\]

In this notation, the first two symmetries of equation (\ref{QV}) (see \cite{TTX, X}) can be written in the following form 
\begin{eqnarray}
&& K^{(1)} = h w- \frac{1}{2} \partial_{u_{1,0}} h =h_{-1} w + \frac{1}{2} \partial_{u_{-1,0}} h_{-1}\ ; \label{K1}\\
&& K^{(2)} = h\,h_{-1} w^2 \left(w_1 +w_{-1} \right)\,.\label{K2}
\end{eqnarray}
In \cite{mr89k:58132,X} it was shown that the above symmetries commute
\begin{equation}\label{comK1K2}
 [ K^{(1)}, K^{(2)}]=D_{K^{(2)}} (K^{(1)})-D_{K^{(1)}} (K^{(2)})=0.
\end{equation}

It is much less known about cosymmetries. For partial differential equations cosymmetries were introduced and 
studied in \cite{wang98}. The simplest cosymmetry for equation (\ref{QV}),  which we have found by direct solution
of equation (\ref{sym_def1}), can be written in the form
\begin{equation}
\omega^{(1)}=\frac{w_1}{Q_{u_{0,0}}}-\frac{w}{Q_{u_{1,0}}}+\frac{\partial_{u_{1,0}} g}
{2  gQ_{u_{1,0}}} +\frac{\partial_{u_{0,0}} h}{2  hQ_{u_{0,0}}}
\label{cosym1}
\end{equation}
It is easy to check that $\omega^{(1)}$ is a cosymmetry. Indeed, using the elimination map one can show that 
\begin{equation}\label{halfomega1}
 \cS^{-1}(Q_{u_{1,0}} \omega^{(1)})+ \omega^{(1)} Q_{u_{0,0}}
=-\cT^{-1}\cS^{-1}(Q_{u_{1,1}} \omega^{(1)})- \cT^{-1}(\omega^{(1)} Q_{u_{0,1}})
 =\phantom{-}\frac{K^{(1)}_{1}}{h}-\frac{K^{(1)}_{-1}}{h_{-1}}.
\end{equation}
This immediately leads to $D_Q^*\omega^{(1)}=0$ and proves that $\omega^{(1)}$ is a cosymmetry of  (\ref{QV})
according to Definition \ref{DefcoSymmetry}.  The next cosymmetry $\omega^{(2)}$ (cf. (\ref{cosym2})) 
looks considerably more complicated and it will be derived in the next Section.                                                                                                                       

Symmetries $K^{(1)},K^{(2)}$ are elements of $\cF_{\bs}$, they do not depend on variables $u_{k,p}$ with $p\ne 0$ and thus it is relatively easy to find them starting from the Definition \ref{DefSymmetry}. Cosymmetries $\omega^{(1)},\omega^{(2)}$ are elements of $\cF_Q$ which cannot be reduced to elements of $\cF_{\bs}$ or $\cF_{\bt}$ by the elimination map. 

Equation (\ref{QV}) has an infinite hierarchy of commuting symmetries as well as an infinite hierarchy of cosymmetries.
In the next sections we shall show that the hierarchy of symmetries can be generated by successive applications of
 a recursion operator to the {\em seed} symmetries $K^{(1)},K^{(2)}$, while the hierarchy of cosymmetries can be obtained 
by applying successively a co-recursion operator to the {\em seed} cosymmetries  $\omega^{(1)},\omega^{(2)}$.

\section{Recursion and co--recursion operators of difference equations}\label{Co-recursion}

In this section we prove that equation (\ref{QV}) possesses weakly nonlocal pseudo-difference recursion and co-recursion operators. 

We remind that difference operators are finite sums of the form 
\begin{equation}
A = a_n\cS^n+a_{n-1}\cS^{n-1}+\cdots +a_m\cS^m,\qquad a_k\in\cF_Q \ .
\label{Adiff}
\end{equation}
They form a non-commutative ring $\cF_Q[\cS]$ of polynomials in $\cS$ with coefficients from $\cF_Q$. Difference operators we also call local operators since  the action $A:a\mapsto A(a)\in\cF_Q$ of a difference operator $A$ is naturally defined  for any element  $a\in\cF_Q$.

For our theory the operator of a finite difference $\Delta=\cS-1$ is of particular importance. This is a 
difference analogue of the derivation $D_x$ in a differential field in the continuous case. 
The corresponding inverse pseudo-difference operator $\Delta^{-1}$ is the analogue of  $D_x^{-1}$ or integration. 
The action of the operator $\Delta^{-1}$ is not defined for all elements of $\cF_Q$, but only on the elements of 
image space ${\rm Im}\,\Delta=\Delta\cF_Q$.  In general the kernel space ${\rm Ker}\,\Delta \subset\cF_Q$ could be 
rather nontrivial (see discussion in \cite{mwx1}), but in the case of equation (\ref{QV}), where $Q$ is an irreducible 
polynomial satisfying (\ref{affinelinear}), one can show that ${\rm Ker}\,\Delta=\bbbc$. Thus for $a=b_1-b, \ b\in\cF_Q$ we
 have $\Delta^{-1} (a)=b+\alpha,\ \alpha\in\bbbc$. In what follows we shall ignore ``the constant of integration'' $\alpha$, since its effect on results is inessential similar to the continuous case \cite{mr86i:58070}.

We define weakly nonlocal pseudo-difference operators as finite sums of the form
\begin{equation}
  B=B_0+a_1 \Delta^{-1}\circ b_1+a_{2}\Delta^{-1}\circ b_{2}+\cdots +a_m\Delta^{-1}\circ  b_m,\qquad a_k,b_k\in\cF_Q,\ B_0\in \cF_Q[\cS].\label{Bweakly}
\end{equation}
It is a difference analogue of weakly nonlocal pseudo-differential operators which play important r\^ole in the theory of multi-Hamiltonian partial differential equations \cite{MaN01}.
A composition of two  weakly nonlocal pseudo-differential operators is a pseudo-difference operator, but not necessarily weakly nonlocal.  

Formally conjugated difference (\ref{Adiff}) and weakly-non local pseudo-difference (\ref{Bweakly}) operators are defined as
\[
 A^*=\cS^{-n}\circ a_{n}+\cS^{1-n}\circ a_{n-1}+\cdots +\cS^{-m}\circ a_{m},
\]
\[
  B^*=B_0^*-b_1 \Delta^{-1}\cS\circ  a_1-b_{2}\Delta^{-1}\cS\circ  a_{2}-\cdots -b_m\Delta^{-1}\cS\circ  a_m.
\]

The action of a pseudo-difference operator is not defined for all elements of $\cF_Q$. The subset of $\cF_Q$ for 
which the action of $B$ is defined ${\rm Dom}(B)=\{a\in\cF_Q\,|\, B(a)\in\cF_Q\}$ is called the domain of $B$. For example 
${\rm Dom}(\Delta^{-1})={\rm Im}\,\Delta$ and for $A\in\cF_Q[\cS]$ we have ${\rm Dom}(A)=\cF_Q$.

By a recursion operator  of a difference equation (\ref{QV}) we shall understand a pseudo-difference operator $\gR$ such that
\[
\gR:{\rm Dom}(\gR)\cap \gA_Q\mapsto \gA_Q,
\]
where $\gA_Q$ is the linear space of symmetries of this difference equation. In other
words, if the action of $\gR$ on a symmetry $K\in\cF_0$ is defined, i.e. $\gR (K)\in
\cF_Q$, then $\gR (K)$ is a symmetry of the same difference equation. The operator of multiplication by a
constant  is a {\em trivial} recursion operator. It easy to see that a pseudo-difference operator $\gR$ is
a recursion operator for a difference equation $Q=0$ if there exists a pseudo-difference operator $\gP$ such that
$\gR$ and $\gP$ satisfy the following operator equation
\begin{equation}\label{rec}
 D_{Q} \circ \gR=\gP \circ D_{Q}.
\end{equation}
Indeed, if $K$ is a symmetry of this difference equation and $\hat{K}=\gR(K)\in\cF_Q$, then it follows from (\ref{rec}) that 
$D_Q \hat{K}=0$, and thus $\hat{K}$ is also a symmetry. 

The equality in (\ref{rec}) has to be understood in the sense of the field $\cF_Q$, or in the usual sense after application of the elimination map to the coefficients of the pseudo-difference operators. In 
 \cite{mwx1} it was shown that the operator equation (\ref{rec}) is equivalent to two equations
\begin{equation}\label{gP1}
 \gP \circ (Q_{u_{1,0}}\cS+Q_{u_{0,0}})=(Q_{u_{1,0}}\cS+Q_{u_{0,0}}) \circ \gR \ ,
\end{equation}
and
\begin{equation}\label{recursion_eq}
 \gP \circ (Q_{u_{1,1}}\cS+Q_{u_{0,1}})=(Q_{u_{1,1}}\cS+Q_{u_{0,1}}) \circ \cT(\gR) \ .
\end{equation}

We say a pseudo-difference operator $\gW$ is a co--recursion operator of a difference equation $Q=0$ 
if it maps cosymmetries to cosymmetries $\gW:\gO_Q\cap {\rm Dom}\gW\mapsto \gO_Q$. If the difference equation possesses a recursion operator
$\gR$ satisfying equation (\ref{rec}) then $\gW=\gP^*$. Indeed, conjugating  (\ref{rec}) we get
\begin{equation}\label{crec}
 D_{Q}^* \circ \gP^*=\gR^* \circ D_{Q}^*
\end{equation}
and thus if $\omega$ is a cosymmetry and $\hat{\omega}=\gW(\omega)\in{\rm Dom}\gW$, then $D_{Q}^*(\hat{\omega})=0$.

In the rest of this section, we show that with the operator $\gR$ presented in paper \cite{mwx1} there exists (and will be given explicitly) a pseudo-difference operator $\gP$ satisfying equation (\ref{rec}). It implies that $\gR$ is 
a recursion operator of the difference equation (\ref{QV}) and  $\gW=\gP^*$ is a co-recursion operator.

We  rewrite operator $\gR$ in \cite{mwx1} in the form
\begin{eqnarray}
\gR &=& h\, h_{-1}\,w^2 w_{1}^2 \cS^2 + h\, h_{-1}\,w^2 w_{-1}^2\cS^{-2}
+2 K^{(1)} K^{(2)} \left(\frac{1}{h} \cS+\cS^{-1}\frac{1}{h}\right) \nonumber \\
&& - w^2 \left(h_{-1}\, h_{1} w_{1}^2+ h_{-2}\, h\,w_{-1}^2\right)
+\frac{2}{h_{-1}} \left( K^{(1)} K^{(2)}_{-1} + K^{(2)} K^{(1)}_{-1} \right) \nonumber \\
& & +\, 2\,\,K^{(1)}\, \Delta^{-1}\circ
\left(\frac{K^{(2)}_{-1}}{h_{-1}}\,-\frac{K^{(2)}_{1}}{h} \right) 
+ 2\,\,K^{(2)}\, \Delta^{-1}\circ 
\left(\frac{K^{(1)}_{-1}}{h_{-1}}\,-\frac{K^{(1)}_{1}}{h} \right)\, ,
\label{weakR}
\end{eqnarray} 
where $h$, $w$, $K^{(1)}$ and $K^{(2)}$ are given by (\ref{h-polynomials-def1}), (\ref{w}), (\ref{K1}) and (\ref{K2}) respectively. 
It is a weakly non-local pseudo-difference operator
and can be written as the product of the following operators
\begin{eqnarray}
{\cal{H}} &=& \frac{h_{-1}\, h\, h_{1}}{(u_{1,0}-u_{-1,0})^2 (u_{2,0}-u_{0,0})^2}\, {\cal{S}}\,-\,
{\cal{S}}^{-1} \frac{h_{-1}\, h\, h_{1}}{(u_{1,0}-u_{-1,0})^2 (u_{2,0}-u_{0,0})^2}  \nonumber \\
& & \nonumber \\
& & +\, 2\,K^{(1)}\,  \Delta^{-1}{\cal{S}} \circ K^{(2)}\, +\, 2\, K^{(2)}\,
\Delta^{-1}\circ K^{(1)} \,, \label{hamilt-gen}\\
\J&=& \frac{1}{h}\ \cS -\cS^{-1}\ \frac{1}{h} \,, \label{symp-gen}
\end{eqnarray}
that is, $\gR={\cal{H}}\circ \J $. In section \ref{locality}, we are going to show that operator $\cal{H}$ 
is Hamiltonian and operator $\J$ is symplectic.
\begin{The}\label{th1}
Operator $\gR$ given by (\ref{weakR}) is a recursion operator of equation (\ref{QV}).
\end{The}
Before we prove this theorem, we first give a few lemmas.
\begin{Lem}\label{lem1}
If a pseudo-difference operator
\begin{eqnarray*}
A= a^{(2)} \cS^2+ a^{(1)} \cS+ a^{(0)} + a^{(-1)} \cS^{-1}+ a^{(-2)} \cS^{-2}+ b_l^{(1)} \Delta^{-1} b_r^{(1)}
+ b_l^{(2)} \Delta^{-1} b_r^{(2)}
\end{eqnarray*}
can be factorised as the product of the following two operators 
\begin{eqnarray*}
C&=& c^{(1)} \cS+ c^{(0)} + c^{(-1)} \cS^{-1}+ c^{(-2)} \cS^{-2}+ b_l^{(1)} \Delta^{-1} d_r^{(1)}
+ b_l^{(2)} \Delta^{-1} d_r^{(2)}, \\ 
E&=& e^{(1)} \cS+ e^{(0)},
\end{eqnarray*}
that is, $A=C\circ E$, then the coefficients of the operator $C$  satisfy the following conditions
\begin{eqnarray}
&&a^{(2)} =c^{(1)} e_1^{(1)} , \qquad\qquad  a^{(-2)}=c^{(-2)}  e_{-2}^{(0)},\label{apm2}\\
&& a^{(1)}=c^{(0)} e^{(1)}+c^{(1)}  e_1^{(0)},\qquad  a^{(-1)}=c^{(-2)} e_{-2}^{(1)} + c^{(-1)}  e_{-1}^{(0)},
\label{apm1}\\
&& b_r^{(1)}=(d_r^{(1)} e^{(1)})_{-1}+d_r^{(1)}  e^{(0)},\label{br1}\\
&& b_r^{(2)}=(d_r^{(2)} e^{(1)})_{-1}+ d_r^{(2)}  e^{(0)},\label{br2}\\ 
&& a^{(0)}= c^{(-1)} e_{-1}^{(1)}  + b_l^{(1)}  (d_r^{(1)} e^{(1)})_{-1}+ b_l^{(2)}  (d_r^{(2)} e^{(1)})_{-1}+ c^{(0)}  e^{(0)}. \label{constr}
\end{eqnarray}
\end{Lem}
{\bf Proof}. Expanding the composition of operators $C$ and $E$ and using the identity $\Delta^{-1}\cS=1+\Delta^{-1}$
we get:
\begin{eqnarray*}
C\circ E&=&c^{(1)} e_1^{(1)} \cS^2+ \left(c^{(0)} e^{(1)}+c^{(1)}  e_1^{(0)}\right) \cS + \left(c^{(-2)} e_{-2}^{(1)} + c^{(-1)}  e_{-1}^{(0)}\right)\cS^{-1} + c^{(-2)}  e_{-2}^{(0)}\cS^{-2}\\
&+&\left( c^{(-1)} e_{-1}^{(1)}  + b_l^{(1)}  (d_r^{(1)} e^{(1)})_{-1}+ b_l^{(2)}  (d_r^{(2)} e^{(1)})_{-1}+ c^{(0)}  e^{(0)}\right) \\
&+& b_l^{(1)} \Delta^{-1} \left((d_r^{(1)} e^{(1)})_{-1}+d_r^{(1)}  e^{(0)}\right)+ b_l^{(2)} \Delta^{-1} \left((d_r^{(2)} e^{(1)})_{-1}+ d_r^{(2)}  e^{(0)} \right)\ .
\end{eqnarray*}
Comparing the coefficients  of the operator $A$ with the above operator we obtain (\ref{apm2})--(\ref{constr}).
\hfill $\blacksquare$

\begin{Lem}\label{lem2}
Let  $Q$ and $\gR$ be given by (\ref{QV}) and (\ref{weakR}) respectively. Then for the constant
\begin{eqnarray}
&&\mu=(a_4-a_3)(a_3 a_1 a_7-a_3 a_5^2-2 a_7 a_2^2+4 a_5 a_2 a_6-2 a_1 a_6^2+a_4 a_1 a_7-a_4 a_5^2)\ , \label{const}
\end{eqnarray}
the operator  $\gR-\mu$ can be factorised over $\cF_Q$ as
\begin{eqnarray}
\gR-\mu &=&\cM \cdot \left( Q_{u_{1,0}} \cS+ Q_{u_{0,0}}\right),\label{fac1}
\end{eqnarray}
where 
\begin{equation}\label{cM}
 \cM= c^{(1)} \cS+ c^{(0)} + c^{(-1)} \cS^{-1}+ c^{(-2)} \cS^{-2}+ 2\,\,K^{(1)} \Delta^{-1} \omega^{(2)}
- 2\,\,K^{(2)} \Delta^{-1} \omega^{(1)}
\end{equation}
with coefficients 
\begin{eqnarray}
&& c^{(1)}= \frac{h\, h_{-1}\,w^2 w_{1}^2}{\cS(Q_{u_{1,0}})}; \qquad  c^{(-2)}= \frac{h\, h_{-1}\,w^2 w_{-1}^2}{\cS^{-2}(Q_{u_{0,0}})};\label{c1m2r}\\
&& c^{(0)}=\frac{1}{Q_{u_{1,0}}} \left(\frac{2 K^{(1)} K^{(2)}}{h}- \frac{h\, h_{-1}\,w^2 w_{1}^2 \cS(Q_{u_{0,0}}) }{\cS(Q_{u_{1,0}})} \right);\label{c0r}\\
&& c^{(-1)}=\frac{1}{\cS^{-1}(Q_{u_{0,0}})} \left(\frac{2 K^{(1)} K^{(2)}}{h_{-1}}- \frac{h\, h_{-1}\,w^2 w_{-1}^2 \cS^{-2}(Q_{u_{1,0}}) }{\cS^{-2}(Q_{u_{0,0}})} \right);\label{cm1r}\\
&&\omega^{(2)}=\frac{K^{(2)}}{h Q_{u_{1,0}}}- \frac{K_1^{(2)}}{h Q_{u_{0,0}}} -\frac{K_1^{(2)}}{h_1 Q_{u_{1,0}}} \frac{\cS(Q_{u_{0,0}})}{\cS(Q_{u_{1,0}})}+\frac{K_1^{(2)} \omega^{(1)} }{ K_1^{(1)}} +\frac{ K_1^{(2)} K^{(1)}}{h K_1^{(1)} Q_{u_{1,0}}}\nonumber\\ 
&&\qquad -\frac{\mu}{2 K_1^{(1)} Q_{u_{1,0}}} + \frac{h_{1}\,w^2 w_{1}^2}{2 K_1^{(1)} } \left(\frac{h\,  \cS^{-1}(Q_{u_{1,0}}) }{Q_{u_{0,0}} \cS^{-1}(Q_{u_{0,0}})}-\frac{h_{-1}}{ Q_{u_{1,0}}} \right)\label{cosym2}\\
&&\qquad +\frac{h\,w_2^2 w_{1}^2}{2 K_1^{(1)} Q_{u_{1,0}}}\left( \frac{h_1\cS(Q_{u_{0,0}})}{\cS(Q_{u_{1,0}})}  \frac{ \cS^2(Q_{u_{0,0}}) }{\cS^2(Q_{u_{1,0}})}-h_2\right) , \nonumber
\end{eqnarray}
and  $\omega^{(1)}$ given by (\ref{cosym1}).  Moreover, $\omega^{(2)}$ is a cosymmetry of equation (\ref{QV}).
\end{Lem}
{\bf Proof}. Using Lemma \ref{lem1}, we can easily determine $c^{(1)}, c^{(0)}, c^{(-1)}$ and $c^{(-2)}$
given by (\ref{c1m2r}--\ref{cm1r}).
It follows from (\ref{halfomega1}) that $\omega^{(1)}$ (\ref{cosym1}) satisfies relation (\ref{br1}). Using condition (\ref{constr}), we get
\begin{eqnarray*}
&&\omega^{(2)}=\frac{1}{2 K_1^{(1)} Q_{u_{1,0}}} \left( - w_1^2 \left(h\, h_{2} w_{2}^2+ h_{-1}\, h_1\,w^2\right)
+\frac{2}{h} \left( K_1^{(1)} K^{(2)} + K_1^{(2)} K^{(1)} \right)-\mu \right)\\
&&\qquad - \frac{1}{2 K_1^{(1)} Q_{u_{0,0}}} \left(\frac{2 K_1^{(1)} K_1^{(2)}}{h} - \frac{h\, h_{1}\,w^2 w_{1}^2 \cS^{-1}(Q_{u_{1,0}}) }{\cS^{-1}(Q_{u_{0,0}})} \right)+\frac{K_1^{(2)} \omega^{(1)} }{ K_1^{(1)}} \\
&&\qquad -\frac{1}{2 K_1^{(1)} Q_{u_{1,0}}} \frac{\cS(Q_{u_{0,0}})}{\cS(Q_{u_{1,0}})} \left(\frac{2 K_1^{(1)} K_1^{(2)}}{h_1}- \frac{h\, h_{1}\,w_2^2 w_{1}^2 \cS^2(Q_{u_{0,0}}) }{\cS^2(Q_{u_{1,0}})} \right) ,
\end{eqnarray*}
which can be rewritten as in (\ref{cosym2}). Substituting the latter into condition (\ref{br2})
\begin{eqnarray*}
\frac{K^{(2)}_{-1}}{h_{-1}}\,-\frac{K^{(2)}_{1}}{h}=\cS^{-1}(Q_{u_{1,0}} \omega^{(2)})+ \omega^{(2)} Q_{u_{0,0}},
\end{eqnarray*}
we obtain the value of $\mu$. Using the elimination map and Definition \ref{DefcoSymmetry} one can directly check that $\omega^{(2)}$ given by (\ref{cosym2}) is a cosymmetry of equation (\ref{QV}). 
\hfill $\blacksquare$

Factorisations (\ref{fac1}) is  quite remarkable. Indeed, the coefficients of the weakly nonlocal pseudo-difference 
operator $\gR-\mu $ are in the field $\cF_\bs$, but the coefficients of both of its factors depend on the variable $u_{0,1}\not\in\cF_\bs$. It is indeed a factorisation over $\cF_Q$.

Similarly, we obtain the following factorisation for operator $\cT(\gR)-\mu$:
\begin{Lem}\label{lem3}
For operator $\gR$ given by (\ref{weakR})and the constant $\mu$ given by (\ref{const}),
the following factorisation holds for all solutions of equation (\ref{QV})
\begin{eqnarray}
\cT(\gR)-\mu &=&\hat{\cM} \cdot \left( Q_{u_{1,1}} \cS+ Q_{u_{0,1}}\right),\label{fac2}
\end{eqnarray}
where 
\begin{equation}\label{hatcM}
 \hat{\cM}= \hat{c}^{(1)} \cS+ \hat{c}^{(0)} + \hat{c}^{(-1)} \cS^{-1}+ \hat{c}^{(-2)} \cS^{-2}- 2\,\,K^{(1)} \Delta^{-1} \omega^{(2)}
+ 2\,\,K^{(2)} \Delta^{-1} \omega^{(1)} .
\end{equation}
Here $\omega^{(1)}$ and $\omega^{(2)}$ are given by (\ref{cosym1}) and (\ref{cosym2}) respectively, and 
\begin{eqnarray}
&& \hat{c}^{(1)}= \frac{\cT(h\, h_{-1}\,w^2 w_{1}^2)}{\cS(Q_{u_{1,1}})}; \qquad  \hat{c}^{(-2)}= \frac{\cT(h\, h_{-1}\,w^2 w_{-1}^2)}{\cS^{-2}(Q_{u_{0,1}})};\label{c1m2}\\
&& \hat{c}^{(0)}=\frac{1}{Q_{u_{1,1}}} \left(\frac{2\cT( K^{(1)} K^{(2)})}{\cT(h)}- \frac{\cT(h\, h_{-1}\,w^2 w_{1}^2) \cS(Q_{u_{0,1}}) }{\cS(Q_{u_{1,1}})} \right);\label{c0}\\
&& \hat{c}^{(-1)}=\frac{1}{\cS^{-1}(Q_{u_{0,1}})} \left(\frac{2 \cT(K^{(1)} K^{(2)})}{\cT(h_{-1})}- \frac{\cT(h\, h_{-1}\,w^2 w_{-1}^2) \cS^{-2}(Q_{u_{1,1}}) }{\cS^{-2}(Q_{u_{0,1}})} \right) .\label{cm1} 
\end{eqnarray}
\end{Lem}

{\bf Proof of Theorem \ref{th1}}. Using the elimination map we can directly show the operator identity
\begin{eqnarray*}
\left( Q_{u_{1,0}} \cS+ Q_{u_{0,0}}\right) \cM= \left( Q_{u_{1,1}} \cS+ Q_{u_{0,1}}\right) \hat{\cM}.
\end{eqnarray*}
We denote the above operator as $\gP$. From Lemmas \ref{lem2} and \ref{lem3}, it follows that
$$ \gP \circ (Q_{u_{1,0}}\cS+Q_{u_{0,0}})=(Q_{u_{1,0}}\cS+Q_{u_{0,0}}) \circ (\gR-\mu).$$
and 
$$ \gP \circ (Q_{u_{1,1}}\cS+Q_{u_{0,1}})=(Q_{u_{1,1}}\cS+Q_{u_{0,1}}) \circ (\cT(\gR)-\mu).$$
Thus we proved formulas (\ref{gP1}) and (\ref{recursion_eq}) for the  pseudo-difference operators $\gR-\mu$ and 
$\gP$. Therefore, operator $\gR-\mu$ is a recursion operator of equation (\ref{QV}). So is operator $\gR$ since $\mu$ is a constant.
\hfill  $\blacksquare$

From the proof of Theorem \ref{th1}, it follows that
\begin{equation}\label{rec-1}
 D_{Q} \circ (\gR-\mu)=\gP \circ D_{Q} 
\end{equation}
for all solutions of equation (\ref{QV}). This leads to the following result:
\begin{The}\label{th2}
Operator $\gW=\gP^*$ is a co-recursion operator of equation (\ref{QV}), where
$$\gP=\left( Q_{u_{1,0}} \cS+ Q_{u_{0,0}}\right) \cM$$ and operator $\cM$ is defined in Lemma \ref{lem2}.
\end{The}

\section{Locality of symmetry hierarchies}\label{locality}
In this section, we prove that recursion operator (\ref{weakR}) is a Nijenhuis operator.
Despite of being weakly non-local, it generates an infinite hierarchy of local commuting symmetry flows. 
The main result of this section is a re-statement of the Theorem proven in the continuous case \cite{wang09} with 
the adjustments to the difference variational complex. 
\subsection{Difference variational complex and Lie derivatives}
Notice that all coefficients of operator $\gR$ are elements in $\cF_{\bs}$. In this section we sketch the difference variational complex over the ring
$\cF_\bs$ \cite{kp85, hm01} in the same spirit as the variational complex \cite{GD79,
mr94j:58081, mr94g:58260, wang98}.  We also adapt the formula of the Lie derivatives along
evolutionary vector fields. 

Recall that $\cF_\bs$ is the field of rational functions of variables $U_\bs=\{u_{n,0}\,|\, n\in\Z\}$.
Without causing any confusion, we denote $u_{n,0}$ by $u_n$for simplicity in what follows.

The field $\cF_\bs$ is also a linear space over $\C$. Let us consider the extended linear space 
$\cL_\bs=\cF_\bs \bigoplus{\rm Span}_\C\{\log a\,|\, a\in\cF_\bs\}$, so that elements of $\cL_\bs$ are linear 
combinations with complex coefficients of elements in $\cF_\bs$ and logarithms of elements in $\cF_\bs$.
For any element $g\in \cL_{\bs}$, we define an equivalence class (or a functional) $\int\! g$
by saying that two elements $g,h\in\cL_{\bs}$ are equivalent if \(g-h\in
 \Delta (\cL_\bs)\).  The space of functionals will be denoted by $\cF_{\bs}'$, it is a linear space over $\C$ and 
it does not inherit a ring or field structure of $\cF_\bs$.

The evolutionary vector fields over the ring $\cF_{\bs}$ 
\[
 X_P=\sum_{n\in\Z} (\cS^n P)
 \frac{\partial }{\partial u_{n}}
\]
form a Lie algebra denoted by $\lieh$. 
The action of any element $P\in \lieh$ on
$\int\!\! g \in \cF_{\bs}'$ can be defined as
\begin{eqnarray}\label{act}
P \circ {\int} g =\int X_P (g)=\int \sum_{n\in\Z} (\cS^n P)
 \frac{\partial g}{\partial u_{n}}=\int D_g[P].
\end{eqnarray}
This is a representation of the Lie algebra $\lieh$. We build up a Lie algebra
complex associated with it. This complex is called the difference variational complex.
Here we give the first few steps.

We denote the space of functional $n$-forms by $\Omega^n$ starting with
$\Omega^0=\cF_{\bs}'$. We now consider the space $\Omega^1$.  For any vertical 1-form
on the ring $\cF_{\bs}$, i.e., $\omega=\sum_{n} h^{(n)} \d u_{n}$, there is a
natural non-degenerate pairing with an element $P\in \lieh$:
\begin{eqnarray}\label{pairing}
<\omega, \ P>=\int \sum_{n\in\Z} h^{(n)} \cS^n P = \int \left( \sum_{n\in\Z} \cS^{-n} h^{(n)}
\right) \  P \ .
\end{eqnarray}
Thus any element of $\Omega^1$ is completely defined by $\xi\ \d u_{0}=\sum_{n}
\cS^{-n} h^{(n)} \d u_{0}$. We simply say $\xi\in \Omega^1$.

The pairing (\ref{pairing}) allows us to give the definition of (formal) adjoint
operators to linear (pseudo-) difference operators \cite{kp85, mr94g:58260}.
\begin{Def} Given a linear operator ${\cal A}: \lieh \rightarrow  \Omega^1$,
we call the operator ${\cal A}^{\star}: \lieh \rightarrow \Omega^1$ the adjoint
operator of ${\cal A}$ if $<{\cal A} P_1, \ P_2>=<{\cal A}^{\star} P_2, \ P_1>$,
where $P_i\in \lieh$ for $i=1,\ 2$.
\end{Def}
Similarly, we can define the adjoint operator for an operator mapping from $\Omega^1$
to $\lieh$, from $\lieh$ to $\lieh$ or from $\Omega^1$ to $\Omega^1$.

The difference variational derivative (Euler operator) of each functional $\int\!\!
g\in \cF_{\bs}'$ denoted by
 $\delta_{u_0} (\int\!\! g) \in \Omega^1$ is defined so that
\begin{eqnarray}\label{euler}
&&<\delta_{u_0} (\int \!g),\ P>=(\d \int\! g) (P)=P \circ {\int} g=\int D_g[P]=\int D_g^{\star}(1) P\nonumber\\
&&=\int \left( \sum_{n\in\Z} \cS^{-n}  \frac{\partial g}{\partial u_{n}} \right) P=
<\sum_{n\in\Z} \cS^{-n}  \frac{\partial g}{\partial u_{n}},\ P> \  ,
\end{eqnarray}
where $\d: \Omega^n\rightarrow \Omega^{n+1}$ is a coboundary operator. Due to the
non-degeneracy of the pairing (\ref{pairing}), we have
$$\delta_{u_0} (\int\! g)=D_g^{\star}(1)=\sum_{n\in\Z} \cS^{-n}  \frac{\partial g}{\partial u_{n}}=
\frac{\partial }{\partial u_{0}}\left(\sum_{n\in\Z} \cS^{-n}  g \right)\in \Omega^1.$$

For any $\xi\in \Omega^1$, it can be shown that $\d \xi=D_{\xi}-D_{\xi}^{\star}$ \cite{kp85,mr94j:58081}. Thus $\d \xi=0$ is equivalent to $D_{\xi} =D_{\xi}^{\star}.$  We say that the $1$-form $\xi$ is closed if $\d \xi=0$. 

We can define the Lie derivative of a given object in the complex along vector field $K\in \lieh$.
It can be expressed explicitly in terms of Fr{\'e}chet derivatives as follows \cite{mr94j:58081}:
\begin{equation}\label{def0}
\begin{array}{l} L_K\!\int\!\! g=\int\! D_{g}[K] \ \ \mbox{for} \ \ \int\!\! g\in\cF_{\bs}';\\
L_K\! h=[K, h] \ \ \mbox{for} \ \ h\in\lieh;\\
L_K\!\xi =D_{\xi}[K] +D_K^{\star}(\xi) \ \ \mbox{for} \ \ \xi\in\Omega^1;\\
L_K\! \gR=D_{\gR}[K] -D_K \gR+\gR D_K \ \ \mbox{for} \ \ \gR: \lieh \rightarrow
\lieh;\\
L_K\! \cH=D_{\cH}[K] -D_K \cH-\cH D_K^{\star}\ \
\mbox{for} \ \  \cH: \Omega^1 \rightarrow \lieh;\\
L_K \!\J=D_{\J}[K] +D_K^{\star} \J+\J D_K\ \ \mbox{for} \ \  \J: \lieh \rightarrow
\Omega^1.\end{array}
\end{equation}

For a given object $\sigma$ and $K\in \lieh$, if $L_K \sigma=0$ we say $\sigma$ is
conserved  (or invariant) along the vector field $K$ and the vector field $K$ is a
symmetry of the object $\sigma$. This complex plays an important role in the study of
differential-difference equations or symmetry flows of difference equations since we
can associate equations (flows) with vector fields. Hence we can define
interesting objects such as conserved densities, symmetries and recursion operators
as invariants along vector fields.

\subsection{Symplectic, Hamiltonian and Nijenhuis operators}
In the context of difference variational complex, we now define symplectic and
Hamiltonian operators. Such definitions are the same as in the continuous case, where
we define them in the context of the complex of variational calculus. All the results
to determine whether a given (pseudo-) differential operator is Hamiltonian or
symplectic are still valid. We shall use them later on without redeveloping them.
\begin{Def} A linear operator ${\cal A}: \lieh \rightarrow \Omega^1$ (or
$ \Omega^1 \rightarrow  \lieh $) is anti-symmetric if $ {\cal A}= -{\cal A}^{\star}$.
\end{Def}

Given an anti-symmetric operator $\J: \lieh\rightarrow \Omega^1$, there is an
anti-symmetric $2$-form associated with it. Namely,
\begin{eqnarray}\label{2fom}
\begin{array}{c}\omega(P,G)=<\J(P),G>=-<\J(G),P>=-\omega(G,P), \quad P,G\in \lieh .
\end{array}
\end{eqnarray}
Here the functional $2$-form $\omega$ has the canonical form \cite{mr94g:58260}
\begin{eqnarray}\label{2fomw}
\omega=\frac{1}{2} \int\!\! \d u_0 \wedge \J \d u_0.
\end{eqnarray}
\begin{Def}
An anti-symmetric operator $\J: \lieh\rightarrow \Omega^1$ is called symplectic if its associated
 $2$-form (\ref{2fomw}) is closed, i.e. $\d\omega=0$.
\end{Def}
\begin{Pro}\label{exs}
For any function $f\in \cF_{\bs}$ depending only on $u_0$ and $u_{1}$, the operator
$f \cS -\cS^{-1} f $ is symplectic.
\end{Pro}
{\bf Proof}. This operator is obviously anti-symmetric and its associated canonical $2$-form is
\begin{eqnarray*}
\omega=\frac{1}{2} \int\!\!  \left( f
\d u_0 \wedge  \d u_1 - \d u_0 \wedge  \cS^{-1} ( f \d u_{0}) \right)=\int\!\!  f
\d u_0 \wedge  \d u_1 \ .
\end{eqnarray*}
Now we have
\begin{eqnarray*}
\d \omega= \int\!\! \left(\frac{\partial f}{\partial u_0} \d u_0 \wedge \d u_0 \wedge
\d u_1 +\frac{\partial f}{\partial u_1} \d u_1 \wedge \d u_0 \wedge \d
u_1 \right)=0 .
\end{eqnarray*}
Therefore $\omega$ is a closed $2$-form and thus the given operator is symplectic. \hfill
$\blacksquare$

It immediately follows that operator $\J$ defined by (\ref{symp-gen}) is symplectic since the difference polynomial
$h$ given by (\ref{h-polynomials-def1}) depending only on $u_{0}$ and $u_{1}$.

For an anti-symmetric operator $\cH: \Omega^1\rightarrow \lieh$, we define a bracket
of two functionals as
\begin{eqnarray}\label{poi}
\begin{array}{l}\left\{  \int\!\! f,\  \int\!\! g\right\}=<\delta_{u_0} (\int\!\! f),\ \cH \delta_{u_0}( \int\!\! g)> .
\end{array}
\end{eqnarray}
\begin{Def}
The operator $\cH$ is Hamiltonian if the bracket defined by (\ref{poi}) is Poisson,
that is,  anti-symmetric and satisfying the Jacobi identity
$$\begin{array}{c}\left\{\left\{  \int\!\!f, \int\!\!g \right\},\
 \int\!\! h \right\}+\left\{\left\{ \int\!\! g,  \int\!\! h \right\},\
 \int\!\!  f \right\}+\left\{\left\{  \int\!\! h,  \int\!\! f \right\},\
\int\!\! g \right\}=0 .\end{array}$$
\end{Def}
The Jacobi identity is abstractly the same as in the continuous case when $\cH$ is a
differential operator. Thus the results to determine whether a given difference
operator is Hamiltonian are still valid. In \cite{mr94g:58260} (see p. $443$), it is
formulated as the vanishing of the functional tri-vector:
\begin{eqnarray}\label{triv}
  X_{\cH (\theta)}(\Theta_{\cH})=0,\quad \mbox{where} \quad
\Theta_{\cH} =\frac{1}{2} \int  \theta \wedge \cH (\theta) \ .
\end{eqnarray}
It is clear that any anti-symmetric constant operator is Hamiltonian. For example,
operators $\cS-\cS^{-1}$, $(\cS+1) (\cS-1)^{-1}$ and $(\cS-1)(\cS+1)^{-1}$ are all
Hamiltonian operators.
\begin{Pro}\label{exh}
Anti-symmetric operator ${\cal{H}}$ defined by (\ref{hamilt-gen})
is a  Hamiltonian operator.
\end{Pro}
{\bf Proof}. According to (\ref{triv}), we check $ X_{{\cal H}(\theta)} ( \Theta_{\cH})=0$, where the associated bi-vector of $\cH$ is
\begin{eqnarray*}
\Theta_{\cH}& =&\frac{1}{2} \int \theta \wedge \cH (\theta) \\
&=&\int  \theta \wedge \left(h_{-1}\, h\, h_{1} w^2 w_1^2 \theta_1+\, 2\, K^{(2)}\,
\Delta^{-1} K^{(1)}  \theta \right)\ .
\end{eqnarray*}
Instead of writing out the full calculation, we only demonstrate
the method by picking out terms of $\theta \wedge \Delta^{-1} K^{(1)}  \theta \wedge \Delta^{-1} K^{(2)}  \theta$.
The relevant terms in $ X_{{\cal H}(\theta)} ( \Theta_{\cH})$ are
\begin{eqnarray}\label{midhh}
\begin{array}{l}\quad
\int \left( \theta \wedge X_{{\cal H}(\theta)} (K^{(2)})\wedge  \Delta^{-1} K^{(1)} \theta+K^{(2)} \theta \wedge  \Delta^{-1} ( X_{{\cal H}(\theta)} (K^{(1)}) \wedge \theta \right)\\
=\int  \theta \wedge X_{{\cal H}(\theta)} (K^{(2)})\wedge  \Delta^{-1} K^{(1)} \theta 
+\int  (\cS^{-1}-1)^{-1} K^{(2)} \theta \wedge  X_{{\cal H}(\theta)} (K^{(1)}) \wedge \theta \\
=\int \theta \wedge X_{{\cal H}(\theta)} (K^{(2)})\wedge  \Delta^{-1} K^{(1)} \theta 
-\int  \Delta^{-1} K^{(2)} \theta \wedge  X_{{\cal H}(\theta)} (K^{(1)}) \wedge \theta \\
=\int  \theta \wedge X_{{\cal H}(\theta)} (K^{(2)})\wedge  \Delta^{-1} K^{(1)} \theta 
+\int  \theta \wedge  X_{{\cal H}(\theta)} (K^{(1)}) \wedge  \Delta^{-1} K^{(2)}\theta \\
=\int  \theta \wedge \sum_{i} \cS^{i}({\cal H}(\theta)) \partial_{u_i} K^{(2)}  \wedge  \Delta^{-1} K^{(1)} \theta 
+\int  \theta \wedge  \sum_{i} \cS^{i}({\cal H}(\theta)) \partial_{u_i} K^{(1)} \wedge  \Delta^{-1} K^{(2)}\theta \ .
\end{array}\end{eqnarray}
Notice that the terms with either $\Delta^{-1} K^{(1)}  \theta$ or $\Delta^{-1} K^{(2)}  \theta$ 
in $\cS^{i}({\cal H}(\theta))$ for $i\in \mathbb{Z}$ equal to
\begin{eqnarray*}
\cS^{i}({\cal H}(\theta)): & 2 K_{i}^{(1)} \Delta^{-1} K^{(2)}\theta+2 K_{i}^{(2)} \Delta^{-1} K^{(1)}\theta.
\end{eqnarray*}
Substituting it into (\ref{midhh}), we obtain that the coefficient of $2 \ \theta \wedge \Delta^{-1} K^{(2)}  \theta \wedge \Delta^{-1} K^{(1)}  \theta$ is:
\begin{eqnarray*}
D_{K^{(2)}} [K^{(1)}]-D_{K^{(1)}} [K^{(2)}]=[K^{(1)},\ K^{(2)}]=0\ .
\end{eqnarray*}
By working out for other terms, we can show that the tri-vector $ X_{\cH (\theta)}(\Theta_{\cH})$ vanishes,
which implies that $\cH$ is a Hamiltonian operator. \hfill $\blacksquare$

The Jacobi identity is a quadratic relation for the operator $\cH$. In general, the
linear combination of two Hamiltonian operators is no longer Hamiltonian. If it is,
we say that these two Hamiltonian operators form a Hamiltonian pair.
Hamiltonian pairs play an important role in the theory of
integrability. They naturally generate Nijenhuis operators.

\begin{Def} A linear operator $\gR: \lieh\rightarrow \lieh$ is called a Nijenhuis
operator if it satisfies
\begin{eqnarray}\label{Nijen}
\begin{array}{l}[\gR P, \gR G]-\gR[\gR P, G]-\gR[P, \gR G]+\gR^2[P, G]=0,
\quad P, G\in \lieh.\end{array}
\end{eqnarray}
\end{Def}

Using formula (\ref{def0}), this identity is equivalent
to
\begin{eqnarray}\label{Nijen1}
\begin{array}{l}L_{\gR G} \gR =\gR L_G \gR ,\quad G\in \lieh.\end{array}
\end{eqnarray}
An equivalent formulation is: $D_{\gR}[\gR P](G)-\gR D_{\gR}[P] (G)$ is symmetric
with respect to $P$ and $G$ \cite{mr84j:58046}.
It can be used to check directly whether a given operator is Nijenhuis or
not \cite{wang09}.

The properties of Nijenhuis operators \cite{mr94j:58081} provide us with the
explanation of how the infinitely many commuting symmetries and conservation laws of
integrable equations arise. In applications, there are nonlocal terms in Nijenhuis
operators. For pseudo-differential operators, a lot of work has
been done to find sufficient conditions for Nijenhuis operators to produce local
objects \cite{ mr1974732, serg5, wang09}.

\begin{Pro}\label{prop3}
Operator $\gR$ defined by (\ref{weakR}) is a Nijenhuis operator.
\end{Pro}

{\bf Proof}.  We need to check that the expression $H:=D_{\gR}[\gR P](G)-\gR D_{\gR}[P] (G)$ is symmetric
with respect to $P$ and $G$. The calculation is straightforward but rather long and not suitable for a presentation in a journal article. Here we show only one step by picking out the
terms in $H$ involving either $G_4$ or $P_4$. We use the notation $P_i=\cS^i P$ and $G_j=\cS ^j G$.
Since $\gR$ is a second order operator, we compute the terms containing either $G_2$ or $P_2$
($P_3$ for nonlocal terms)  
in expression $D_{\gR}[P] (G)$. These terms are
\begin{eqnarray}
&& -2 h\, h_{-1}\,w^3 w_{1}^2 (P_1-P_{-1}) G_2
-2 h\, h_{-1}\,w^2 w_{1}^3 (P_2-P) G_2   +D_h[P]   h_{-1}\,w^2 w_{1}^2  G_{2}\nonumber\\
&&
+ h\, D_{h_{-1}}[P] \,w^2 w_{1}^2  G_{2}-2 K^{(1)} h_{-1} w^2 w_1^2 P_2 G_1
-2 K^{(1)} h w^2 w_1^2 P_2 G_{-1} +2 w^2 h_{-1}\, h_{1} w_{1}^3 P_2 G\nonumber\\
&&  -h_{-1} w^2 w_1^2 \left(\partial_{u_{2}} h_{1} \right) P_2 G-2 K^{(1)}_{-1} h w^2 w_1^2 P_2 G + 2\,\,K^{(1)}\, \Delta^{-1} \left(h_{1} w_1^2 w_2^2 P_3 G \right)\nonumber\\
&& -2 w^2 h h_{-1} w_1^2 P_2 \ \Delta^{-1} \left(K^{(1)}_{-1} G/h_{-1}\,-K^{(1)}_{1}G/h \right) \label{q2p2} .
\end{eqnarray} 
Now the term $h\, h_{-1}\,w^2 w_{1}^2 \cS^2$ in $\gR$ acting on the above expression leads to 
the terms with either $G_4$ or $P_4$ in $\gR D_{\gR}[P] (G)$, denoted by $H^{1,4}$. We obtain
\begin{eqnarray*}
&&H^{1,4}=-2  h_{-1} h h_1 h_2\,w^2 w_{1}^2 w_2^3 w_{3}^2 (P_3-P_{1}) G_4
-2 h_{-1} h h_1 h_2 \,w^2 w_{1}^2 w_2^2 w_{3}^3 (P_4-P_2) G_4 \\
&&\quad  +  h_{-1} h h_1\,w^2 w_{1}^2 w_2^2 w_{3}^2 D_{h_2}[P] \,  G_{4}
+ h_{-1} h h_2 \,w^2 w_{1}^2 w_2^2 w_{3}^2  D_{h_{1}}[P] \, G_{4}-2 h_{-1} h h_1 \,w^2 w_{1}^2 K_2^{(1)} 
 w_2^2 w_3^2 P_4 G_3\\
&& \quad -2h h_{-1} h_2\,w^2 w_{1}^2 K_2^{(1)}  w_2^2 w_3^2 P_4 G_{1} +2 h\, h_{-1} h_1 h_3 \,w^2 w_{1}^2
w_2^2 w_{3}^3 P_4  G_2  - h_{-1}\,h h_1 w^2 w_{1}^2 w_2^2 w_{3}^2 \left(\partial_{u_{4}} h_{3} \right) P_4  G_2
\\
&&\quad-2 h\, h_{-1}\,w^2 w_{1}^2 K^{(1)}_{1} h_2 w_2^2 w_3^2 P_4 G_2+ 2 h\, h_{-1} h_2\,w^2 w_{1}^2 \,\,K_2^{(1)}\,
 w_2^2 w_3^2 P_4 G_1\\
&&\quad -2 h_2 \,w^2 w_{1}^2 w_2^2   w_3^2 P_4 \left( h_1 h_{-1}K^{(1)} G_1-
h_{-1} h  K^{(1)}_{2} G_1 +h h_1 K^{(1)}_{-1} G-h_{-1} h_1 K^{(1)}_{1} G\right)\\
&&\quad =-2 h_{-1} h h_1\,w^2 w_{1}^2 w_2^2 w_{3}^2 \left(h_2 w_2 -\frac{1}{2} \partial_{u_{3}} h_2\right) P_3 G_4-2 h_{-1} h h_1 \,w^2 w_{1}^2 K_2^{(1)} 
 w_2^2 w_3^2 P_4 G_3\\
&&\quad
+2 h\, h_{-1} h_1 \,w^2 w_{1}^2 w_2^2 w_{3}^2 \left( h_2 w_{3}+\frac{1}{2} \partial_{u_{2}} h_2\right) P_2 G_4+2 h\, h_{-1} h_1  \,w^2 w_{1}^2
w_2^2 w_{3}^2 \left( h_3 w_{3}-\frac{1}{2} \partial_{u_{4}} h_3\right) P_4  G_2 
\\
&&\quad +2  h_{-1} h  h_2\,w^2 w_{1}^2 w_2^2 w_{3}^2 \left( h_1 w_{2}+\frac{1}{2} \partial_{u_{1}} h_1\right) P_{1} G_4+ 2 h\, h_{-1} h_2\,w^2 w_{1}^2 \,\,K_2^{(1)}\,
 w_2^2 w_3^2 P_4 G_1
 \\
&&\quad+ h_{-1} h h_2 \,w^2 w_{1}^2 w_2^2 w_{3}^2 \left( \partial_{u_{2}} h_1\right) \,P_2  G_{4}
-2 h\, h_{-1} h_2 \,w^2 w_{1}^2  w_2^2 w_3^2 \left(h_1 w_1-\frac{1}{2} \partial_{u_{2}} h_1 \right)P_4 G_2\\
&&\quad -2 h_2 \,w^2 w_{1}^2 w_2^2   w_3^2 P_4 \left( h_1 h_{-1}K^{(1)} G_1 +h h_1 K^{(1)}_{-1} G-h_{-1} h_1 K^{(1)}_{1} G\right) \ .
\end{eqnarray*}
We then collect the terms with either $G_4$ and $P_4$ in $ D_{\gR}[\gR P] (G)$, denoted by $H^{2,4}$. These 
terms can be obtained from the terms containing $P_2$ in (\ref{q2p2}). Simply replacing $P_2$ by $\cS^2 (\gR P)$, we get
\begin{eqnarray*}
&&-2 h\, h_{-1}\,w^2 w_{1}^3 (\gR P)_2 G_2  -2 K^{(1)} h_{-1} w^2 w_1^2 (\gR P)_2 G_1 -2 K^{(1)} h w^2 w_1^2 (\gR P)_2 G_{-1}\\
&& \quad
+2 w^2 h_{-1}\, h_{1} w_{1}^3 (\gR P)_2 G - w^2 h_{-1}\, \left(\partial_{u_{2}} h_{1} \right) (\gR P)_2 w_{1}^2 G
-2 K^{(1)}_{-1} h w^2 w_1^2 (\gR P)_2 G \\
&& \quad + 2\,\,K^{(1)}\, \Delta^{-1}
\left(h_{1} w_1^2 w_2^2 (\gR P)_3 G \right)\ .
\end{eqnarray*}
It follows that
\begin{eqnarray*}
&&H^{2,4}=-2 h\, h_{-1} h_1 h_2 \,w^2 w_{1}^3 w_2^2 w_3^2 P_4 G_2 
 -2 K^{(1)} h_{-1} h_1 h_2 \,w^2 w_{1}^2 w_2^2 w_3^2 P_4 G_1 
\\&& \quad
+2 w^2 h_{-1}\, h_{1}^2 h_2 w_{1}^3 w_2^2 w_3^2 P_4 G -  h_{-1}\, h_1 h_2 w^2 w_1^2 w_2^2 w_3^2 \left(\partial_{u_{2}} h_{1}\right)  P_4 G
-2 K^{(1)}_{-1} h h_1 h_2 w^2 w_1^2  w_2^2 w_3^2 P_4 G
\\&& \quad=
-2 h\, h_{-1} h_1 h_2 \,w^2 w_{1}^3 w_2^2 w_3^2 P_4 G_2 
 -2 K^{(1)} h_{-1} h_1 h_2 \,w^2 w_{1}^2 w_2^2 w_3^2 P_4 G_1 
\\&& \quad
+2 w^2 h_{-1}\, h_{1} h_2 w_{1}^2 w_2^2 w_3^2 K^{(1)}_{1} P_4 G 
-2 K^{(1)}_{-1} h h_1 h_2 w^2 w_1^2  w_2^2 w_3^2 P_4 G \ .
\end{eqnarray*}
Using relation (\ref{K1}), it is easy to see that $H^{1,4}-H^{2,4}$ is symmetric with respect to $P$ and $G$. 
In a similar way, one can check the symmetric property of the remaining terms. \hfill $\blacksquare$
\subsection{Locality of symmetries generated by recursion operators}
The sufficient conditions for Nijenhuis pseudo-differential operators to produce local objects 
formulated in \cite{serg5, mr1974732, wang09} are valid for pseudo-difference operators over 
the field $\cF_\bs$. In a recent paper \cite{wang09}, we proved that 
Nijenhuis operators, which are the product of weakly nonlocal Hamiltonian 
and symplectic operators \cite{MaN01}, generate hierarchies of commuting local symmetries and
conserved densities in involution under some easily verified
conditions. To be self-contained, we restate this result in \cite{wang09} for Hamiltonian operator $\cH$ (\ref{hamilt-gen})
and  symplectic operator (\ref{symp-gen}).
\begin{quotation}
\noindent Consider a Hamiltonian operator $\cH$ of the form (\ref{hamilt-gen}) and a symplectic
operator (\ref{symp-gen}) such that $\gR=\cH\cdot \J$ is a Nijenhuis operator (Proposition \ref{prop3}). Assume that $$L_{K^{(1)}} K^{(2)}
=L_{K^{(1)}} \J=L_{K^{(1)}}\cH =L_{K^{(2)}} \J=L_{K^{(2)}}\cH =0.$$ Then the vector fields $p^{(i,j)}=(\cH \J)^j K^{(i)} \in \lieh$ 
commute and $\omega^{(i,j)}=\J p^{(i,j)}\in \Omega^1$ are
closed $1$-forms for $i=1,2$ and $j=0, 1, 2 , \cdots $.
\end{quotation}
\begin{The}\label{th3} For operator $\gR$ defined by (\ref{weakR}), all $\gR^j K^{(i)}$ commute
and all $1$-forms $\J \gR^j K^{(i)}$ are closed for $i=1,2$ and $j=0, 1, 2 , \cdots $.
\end{The}

{\bf Proof}. We know from (\ref{comK1K2}) that $K^{(1)}$ and $K^{(2)}$ are commuting generalised symmetries. 
Therefore, we need to verify that $L_{K^{(1)}} \J=L_{K^{(1)}}\cH=0$ while we shall skip the proof of the same properties 
for $K^{(2)}$. 

We can write  anti-symmetric operators $\J$ and $\cH$ as $\J=Y-Y^{\star}$ with operator 
$Y=\frac{1}{h} \cS$, and $\cH=Z-Z^{\star}$ with
$$Z=h_{-1}\, h\, h_{1} w^2 w_1^2\, {\cal{S}} +\, 2\, K^{(2)}\,
\Delta^{-1}\circ K^{(1)} \,.$$
We denote operator $D_{Y}[K^{(1)} ] +D_{K^{(1)}} ^{\star} Y+Y D_{K^{(1)}}$ by $F$. It follows from (\ref{def0})
that $L_{K^{(1)}} \J=F-F^{\star}$. Similarly, denoting operator $D_{Z}[K^{(1)}] -D_{K^{(1)}} Z-Z D_{K^{(1)}}^{\star}$
by $G$, we have that $L_{K^{(1)}}\cH=G-G^{\star}$.

We now compute operators $F$ and $G$. 
The Fr{\'e}chet derivative of $K^{(1)}$ and its adjoined operator are given by
\begin{eqnarray*}\label{fk1}
D_{K^{(1)}}=-h_{-1} w^2\ \cS+ \partial_{u_{0}} K^{(1)}+h w^2\ \cS^{-1} \quad \mbox{and} \quad 
D_{K^{(1)}}^{\star}=h_{1} w_1^2\ \cS+ \partial_{u_{0}} K^{(1)}-h_{-2} w_{-1}^2\ \cS^{-1} .
\end{eqnarray*}
Substituting them into the expression for $F$, we obtain 
\begin{eqnarray*}
F&=&-h^{-2} (K^{(1)}_1 \partial_{u_{1}} h +K^{(1)} \partial_{u_0} h )  \cS
+ h^{-1} \partial_{u_{0}} K^{(1)} \cS-h_{-1}^{-1} h_{-2} w_{-1}^2 
+ h^{-1}\partial_{u_{1}} K_1^{(1)}\cS +h^{-1} h_1 w_1^2
\nonumber\\
&=&-h_{-1}^{-1} h_{-2} w_{-1}^2+h^{-1} h_1 w_1^2 , \label{k1If}
\end{eqnarray*}
where we also used $K^{(1)} = h w- \frac{1}{2} \partial_{u_{1}} h$
and $K^{(1)}_1=h w_1 + \frac{1}{2} \partial_{u_{0}} h$ from (\ref{K1}). Hence
\begin{eqnarray}
&&L_{K^{(1)}} \J=F-F^{\star}=0 . \label{k1I}
\end{eqnarray}
By a similar calculation, we find operator $G$ being a symmetric difference operator. It leads to
\begin{eqnarray}
&&L_{K^{(1)}} \cH=G-G^{\star}=0 . \label{k1H}
\end{eqnarray}
Thus we proved the statement. \hfill $\blacksquare$
\subsection{Yamilov's discretisation of the Krichever-Novikov equation}
It follows from (\ref{k1I}) and (\ref{k1H}) that 
$$L_{K^{(1)}} \gR= (L_{K^{(1)}} \cH) \J+\cH L_{K^{(1)}} \J=0\ .$$
This implies that operator $\gR$ is a recursion operator of differential-difference equation $u_{t_1}=K^{(1)}$.
Substituting $Q$ (\ref{QV}) into  $h$ defined by (\ref{h-polynomials-def1}) and then $h$ into  (\ref{K1}) 
we can rewrite the latter equation in the form 
\begin{equation}
 u_{t_1}=\frac{R(u_1,u, u_{-1})}{u_1-u_{-1}}, \label{yamV4}
\end{equation}
 where $R$ is a polynomial 
$$
R(u,v,q)=( \alpha v^2+ 2 \beta v+\gamma) u q + ( \beta v^2+ \lambda v +\delta) (u +q) + \gamma v^2 +2 \delta v +\epsilon
$$
and 
\begin{eqnarray}\label{para}
\begin{array}{lll} \alpha=2 (a_2^2-a_1 a_3); & \beta=a_2 a_5+a_2 a_4-a_2 a_3 -a_1 a_6; & \gamma=2 (a_4 a_5- a_2 a_6);\\
\lambda=a_4^2+a_5^2-a_3^2-a_1 a_7; & \delta=a_4 a_6+a_5 a_6-a_3 a_6-a_2 a_7; & \epsilon=2 (a_6^2-a_3 a_7),
 \end{array}
\end{eqnarray}
where $a_i$ are constant parameters for the Viallet equation (\ref{QV}).
Equation (\ref{yamV4}) can be identified as Yamilov's discretization of the Krichever-Novikov equation (YdKN) \cite{Yami1}, cf. equation (V4) when $\nu=0$ 
in \cite{Yami}. Such relations for all the ABS equations and their generalisations introduced in \cite{TTX} was established in \cite{LeviYami}.

It is straightforward to check that $R(u_1,u, u_1)$ is a symmetric and bi-quadratic polynomial and is related to 
$h$ defined by (\ref{h-polynomials-def1}) as follows
\begin{eqnarray*}
 h(u,u_{1}) =\frac{1}{2}R(u_1,u, u_1).
\end{eqnarray*}
Thus we can express $K^{(2)}$ (\ref{K2}),
recursion operator $\gR$ (\ref{weakR}), Hamiltonian operator (\ref{hamilt-gen}) and symplectic operator
(\ref{symp-gen}) in terms of polynomial $R(u,v,q)$.  For example
\begin{equation}
u_{t_2}=K^{(2)}=\frac{1}{4}\frac{R(u_1,u,u_1)R(u,u_{-1},u)}{(u_1-u_{-1})^2}\left(\frac{1}{u_2-u}+\frac{1}{u -u_{-2}}\right)
\label{yamsym}
\end{equation}
is the next member in the hierarchy of commuting symmetries of the YdKN equation (\ref{yamV4}), which was first given in \cite{mr89k:58132}. Higher symmetries 
can be obtained by application of the recursion operator to the seeds (\ref{yamV4}) and (\ref{yamsym}). 
The locality and commutativity of these symmetries are guaranteed by Theorem \ref{th3}. Cosymmetries of equation 
(\ref{yamV4}) coincide with its covariants, i.e. the variational derivatives of the conserved densities. 
Conserved densities can be obtained as residues of the powers of the recursion operator. They coincide with the 
conserved densities for the Viallet equation (\ref{QV}) \cite{mwx1}. Obviously, equations (\ref{yamV4}) and (\ref{yamsym}) 
and every member of the hierarchy are multi-Hamiltonian systems. For example equation (\ref{yamV4}) can be written 
in a Hamiltonian form
\[
 \cI(u_{t_1})=\frac{\delta H_0}{\delta u},
\]
where the symplectic operator is of the form
 \[
  \cI=\frac{2}{R(u_1,u, u_1)}\cS-\cS^{-1}\frac{2}{R(u_1,u, u_1)}
 \]
and the Hamiltonian $H_0=\ln (u_1-u_{-1})-\frac{1}{2}\ln R(u,u_{-1}, u)=-\frac{1}{2}{\rm res}\,\ln \gR$. 
It follows from Theorem \ref{th3} that $\cI_n=\cI\gR^n$ are symplectic operators. The compatibility of these 
Hamiltonian structures follows from  the Nijenhuis property of the recursion operator. 

\section*{Acknowledgments}
AVM and PX would like to thank the University of Kent for its hospitality during their visits.
JPW is grateful to the University of Kent for granting the study leave.
PX is supported by the {\emph{Newton International Fellowship}} grant NF082473 entitled
``Symmetries and integrability of lattice equations and related partial differential equations''.

\end{document}